# Machine Learning and Quantum Intelligence for Health Data Scenarios

Sanjeev Naguleswaran, QSPectral Systems

## Introduction

The advent of quantum computing has opened new possibilities in data science, offering unique capabilities for addressing complex, data-intensive problems. Traditional machine learning (ML) algorithms often face challenges in high-dimensional or limited-quality datasets, which are common in healthcare. Quantum Machine Learning (QML) leverages quantum properties, such as superposition and entanglement, to enhance pattern recognition and classification, potentially surpassing classical approaches. This paper explores QML's application in healthcare, focusing on quantum kernel methods and hybrid quantum-classical networks for heart disease prediction and COVID-19 detection, assessing their feasibility and performance.

## Background

Quantum computing exploits quantum mechanical principles like superposition and entanglement to solve problems that are difficult for classical systems. Superposition allows quantum systems to exist in a combination of multiple states, and entanglement provides correlation between particles, enabling faster computations. These features, combined with recent advancements in quantum hardware, form the foundation for QML. This paper focuses on how quantum kernel methods can be used in both Support Vector Machines (SVM) and hybrid quantum-classical Neural Networks to extract patterns in data.

Previous work investigated one such challenging domain in predicting start-up success where data scarcity and quality limit classical methods [1]. This earlier work explored how Quantum Machine Learning (QML) can address these issues by creating quantum feature spaces, which offer higher abstraction than classical methods. Additionally, the advent of large-scale quantum processors promises faster processing, making QML a viable tool for data-scarce environments.

Quantum algorithms based on quantum kernel methods have also been investigated in [2]. Previous investigations have also reported advantages obtained due to quantum algorithms leveraging inherent efficiencies in computing over certain group structures. For example, a data-dependent projected quantum kernel was shown to provide a significant advantage over classical kernels in a toy problem [3].

### Quantum Machine Learning Overview

Support Vector Machines are powerful tools that classify data by maximising separation between classes in high-dimensional space. By replacing classical kernels with quantum kernels, we leverage quantum feature spaces that are not attainable by classical systems. These quantum kernels play a critical role in achieving a quantum advantage, especially in scenarios involving limited or poor-quality data. Additionally, we apply quantum kernels as pre-processing layers in Neural Networks (NN) to enhance feature extraction, enabling deeper learning architectures to benefit from quantum-enhanced feature spaces.

In particular, we discuss the investigation and development of quantum kernel functions that leverage quantum-enhanced feature spaces for favourable representations of real-world data; and

## Methodology

We consider two medical applications:

- Predicting heart disease (tabular data)
    - The Behavioural Risk Factor Surveillance System (BRFSS) is a health-related telephone survey that is collected annually by the CDC. This dataset contains 253,680 survey responses from cleaned BRFSS 2015 to be used primarily for the binary classification of heart disease. There is a strong class imbalance in this dataset. 229,787 respondents do not have/have not had heart disease while 23,893 have had heart disease [4].
    - First, a classical Support Vector Machine (SVM) based Machine Learning method was developed and compared to other classical algorithms such as Random Forest.
    - After the performance parity of the SVM was established a Quantum kernel method was developed using the Pennylane Quantum Application Programming Interface (API) [5] - this kernel was substituted for the classical kernel.

- Covid Lung image data obtained from [6]
    - The dataset is organised into two folders (train and test), and both train and test contain two subfolders (COVID-19, NORMAL). The original dataset contains a total of 6432 x-ray images, and test data comprised approximately 20% of the total images.
    - The results in this paper were produced using a reduced dataset of 181 train images and 46 test images. This reduction was necessary due to the use of quantum simulators that are computationally complex on classical computers.

In the first case, we first sampled the data to construct a reduced and balanced dataset for experimentation. Once classical SVM and NN model architectures are established, we substitute quantum kernels to test for potential advantages. Using the Pennylane API, we developed quantum versions of the classical SVM and integrated quantum processing into a hybrid quantum-classical NN architecture. Our experiments involve medical applications like heart disease and COVID-19 lung image classification.

Support Vector Machines (SVMs) are a class of supervised learning algorithms that can be used for classification or regression tasks. They are based on the idea of finding a hyperplane in high-dimensional space that maximally separates different classes. The SVM algorithm separates the classes (such as heart disease vs no heart disease) by drawing a hyperplane between the data points [7]. In cases where the data cannot be linearly separable because the algorithm is based on kernel methods, a mathematical concept known as the "kernel trick" is of particular interest in extending the method to leverage Quantum Computing. The kernel trick works by applying a non-linear transformation to the data before it is fed into the algorithm. This transformation projects the data into a higher-dimensional space, where it may become linearly separable. The algorithm can then be applied to the transformed data, and the resulting model can be used to make predictions on the original data [7].

The kernel trick allows SVMs to handle non-linear classification tasks by implicitly mapping the data into a higher-dimensional space using a kernel function. This allows the algorithm to find a non-linear decision boundary that can accurately separate the classes in the data. However, when the feature space is large the method encounters limitations due to the computational complexity of estimating the kernel functions. The use of a quantum feature space that can only be calculated efficiently on a quantum computer potentially allows for deriving a quantum advantage. Previous work on quantum kernel methods used in ML is described in. Further, an important approach in the search for practical quantum advantage in machine learning is to identify quantum kernels for learning problems that have underlying structure in the data. Examples of learning problems for data with group structure have been identified and a class of kernels related to covariant quantum measurements were constructed [8]. In addition, a data-dependent projected quantum kernel was shown to provide significant advantage over classical kernels.

Further, we consider Quanvolutional Networks for image processing. A Quanvolutional Neural Network (QNN) is a hybrid architecture that integrates quantum computing principles into classical convolutional neural networks (CNNs) to enhance feature extraction. In this model, a quantum processing layer (often referred to as a quantum convolution or "quanvolution") is used as a pre-processing step before the classical convolution layers. The quantum layer processes small patches of the input data, transforming them into quantum states, and outputs a classically interpretable feature map. This method allows the network to capture more complex patterns in data, potentially leading to a performance boost in certain tasks, especially those involving highly complex or non-linear data distributions [9].

## Results and Implications

The ML performance on the heart disease dataset was first benchmarked using a classical SVM algorithm. The methods' performance can be evaluated using the accuracy report and visualising the confusion matrix.

A reduced dataset was used to conduct these experiments due to the computational load of simulating a quantum process on a classical computer. The random selection of 1000 training data points allowed for mitigating the class imbalance by selecting an equal number of data points from each class. The relatively large test set also consisted of 1000 data points for better generalisation and detection of edge cases.

We see that even with a reduced dataset, the quantum support vector machine is comparable to the classical version with a slightly better. This means that it is better at identifying patients with heart disease. We expect the accuracy to be much better when using the full dataset and the full dataset as a test of the classical case with synthetic data balancing yielded an accuracy of 90%.

|  | Precision | Recall | F1-Score | Support |
|---|---|---|---|---|
| No Heart Disease | 0.79 | 0.72 | 0.75 | 507 |
| Heart Disease | 0.74 | 0.80 | 0.77 | 493 |
|  |  |  |  |  |
| accuracy |  |  | 0.76 | 1000 |
| macro avg. | 0.76 | 0.76 | 0.76 | 1000 |
| weighted avg. | 0.76 | 0.76 | 0.76 | 1000 |

*Table 1 Accuracy Report for Classical SVM*

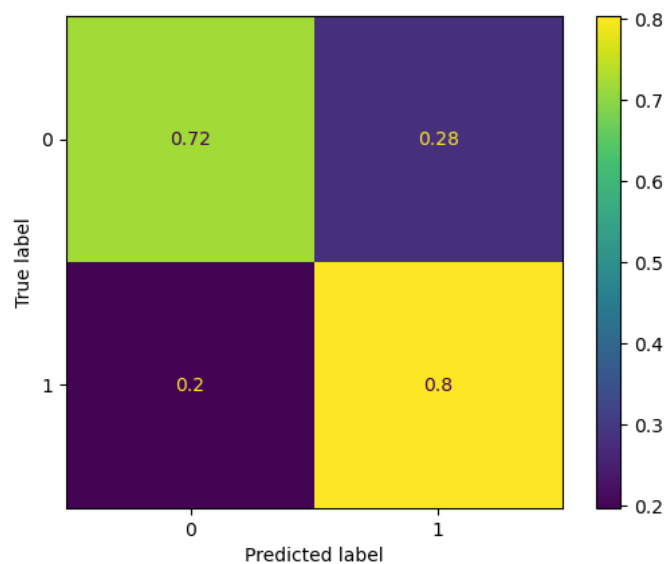

*Figure 1 Confusion matrix for SVM showing the correct and misrepresentation of classes.*

A quantum kernel was then developed and substituted for the classical kernel in the algorithm with the following outcome.

|  | Precision | Recall | F1-Score | Support |
|---|---|---|---|---|
| Class 0 | 0.79 | 0.68 | 0.73 | 507 |
| Class 1 | 0.71 | 0.81 | 0.76 | 493 |
|  |  |  |  |  |
| accuracy |  |  | 0.74 | 1000 |
| macro avg. | 0.75 | 0.75 | 0.74 | 1000 |
| weighted avg. | 0.75 | 0.74 | 0.74 | 1000 |

Table 2 Accuracy Report for Quantum SVM

That performance is comparable to classical SVM. The outcome is also represented in the confusion matrix shown below.

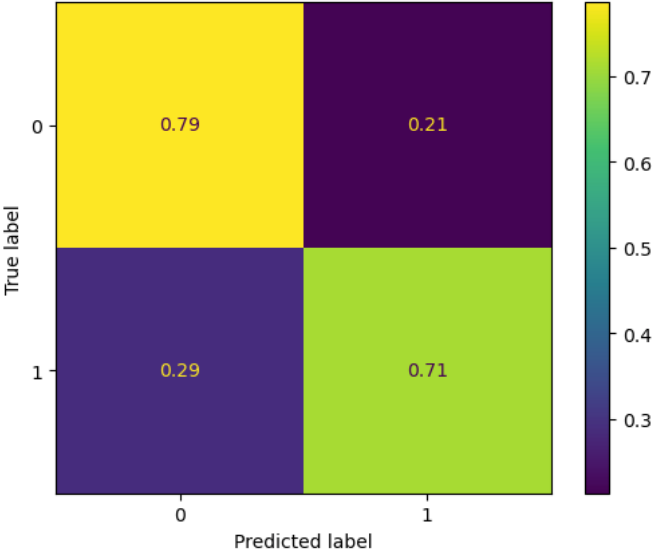

Figure 2 Confusion matrix for Quantum SVM

The above results were obtained with minimal hyperparameter tuning in both cases. We expect the results to improve with tuning and expect the quantum SVM to have different parameter settings to the classical case.

## Hybrid Neural-Network

A further experiment where a quantum convolutional layer was added to a classical Neural Network as a pre-processing layer was conducted. Image identification of COVID-affected vs. Normal lungs was performed. A comparison of performance between a network with a Quantum pre-processing layer and a regular NN architecture was obtained.

A simple, fully connected NN model with 10 output nodes and a SoftMax activation function is used to evaluate the effect of quantum pre-processing. The model was compiled with a stochastic gradient descent optimiser and a cross-entropy loss function.

The following comparisons in accuracy and loss with and without t quantum pre-processing are shown below:

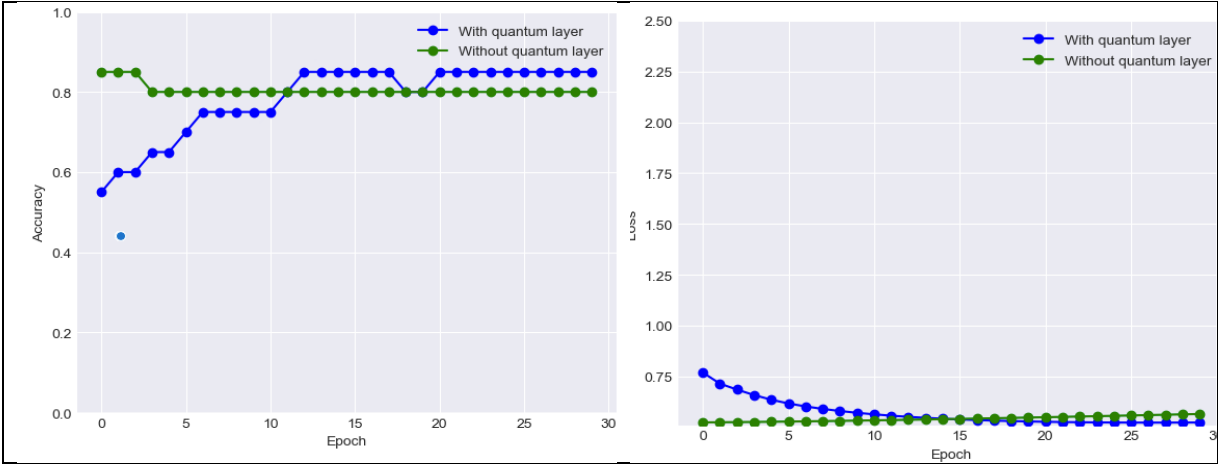

*Figure 3 Comparison of Accuracy and Loss after including quantum pre-processing*

It was shown that when the quantum convolutional method was used as a pre-processor in a hybrid quantum-classical Neural Network there was an advantage after 10 epochs.

## Conclusion

In this study, we demonstrated the potential of Quantum Machine Learning (QML) in medical data analysis by applying quantum kernel methods and hybrid quantum-classical neural networks. Our experiments in heart disease prediction and COVID-19 detection revealed performance parity with classical approaches.

Additionally, when a quantum convolutional method was used as a pre-processing layer in a hybrid quantum-classical Neural Network, an advantage over a purely classical network was observed with the hybrid Neural Network. The Neural Network reported here had a simple architecture. However, the performance improvement with overall higher accuracy was observed even when applying the pre-processing layer to deep learning architecture such as a Convolutional Neural Network (CNN).

These findings underscore QML's feasibility and future promise in real-world health data applications. Further optimisation, including tailored data structures and hyperparameter tuning, could enhance these advantages, marking a significant step towards practical quantum-enhanced healthcare solutions.